\documentclass[twocolumn,showpacs,prl,amsmath,amssymb]{revtex4}
\usepackage{graphicx}
\usepackage{bm}
\begin{document}

\title{Chirality in Quantum Computation with Spin Cluster Qubits}
\author{V.W. Scarola, K. Park and S. Das Sarma}
\affiliation{Condensed Matter Theory Center, 
Department of Physics, University of Maryland,
College Park, MD 20742-4111}

\begin{abstract}
We study corrections to the Heisenberg interaction between 
several lateral, single-electron quantum dots.  We show, 
using exact diagonalization, that three-body chiral terms
couple triangular configurations to external sources of flux rather
strongly.  The chiral corrections impact single qubit encodings 
utilizing loops of three or more Heisenberg coupled quantum dots.
\end{abstract}
\maketitle

Qubits based on electron spin in semiconductor structures 
play a central role in solid state quantum information processing 
\cite{Loss,Hu,Kane,Vrijen}.  In particular, the extremely long 
\cite{DeSousa} spin coherence times $(T_2\gg \mu
\text{s})$ coupled with the enormous lithographic and
fabrication scalability advantages of existing semiconductor
technology have led to a large number of proposed quantum computer 
architectures using electron spin qubits in semiconductor
nanostructures.  Most of these proposals \cite{Loss,Hu,Kane,Vrijen} 
utilize the Heisenberg exchange coupling (``the exchange
gate'' \cite{Loss}) between two electrons localized in individual 
quantum dots \cite{Loss} or in shallow donor states 
(e.g., P atoms in Silicon \cite{Kane}) in order to carry out the 
required two-qubit operations.  The original proposals involve control, 
tuning, and manipulation of one-qubit and two-qubit gates through 
externally applied, local magnetic and electric field pulses
respectively, but it was soon realized that controlling {\em single} 
localized electron spins through local magnetic field (and/or
microwave) pulses is problematic \cite{Hu2,DiVincenzo,Benjamin}, 
perhaps even prohibitively so, and therefore a number of 
interesting theoretical proposals have recently been made 
\cite{Levy,Meier,Benjamin,DiVincenzo} for using a 
cluster of spins (``spin cluster'') 
rather than a single spin, as the suitable qubit unit in solid state
spin based quantum computing architectures.  In particular, some 
elegant, recent theoretical literature
\cite{Bacon,Viola2,Kempe,Levy,Meier,Benjamin,DiVincenzo} has 
convincingly demonstrated as a matter of principle the non-essential
and therefore disposable character of single-spin qubit operations 
altogether, showing that the Heisenberg exchange coupling between 
electron spins by itself can actually
carry out quantum computation in spin-based architectures, 
provided each {\em logical} qubit is encoded in a number of 
{\em physical } spins (i.e., spin cluster encoded qubits).  

One 
important perceived advantage of such collective 
spin based logical qubits (avoiding single electron spin qubits
altogether) is that the system can be placed in a 
decoherence free subsystem or subspace 
\cite{Whaley,Viola}
(where the encoding scheme strongly suppresses 
decoherence of single logical qubits) making such 
encoded spin cluster qubits highly desirable from
the perspective of scalable and coherent fault-tolerant quantum 
computation.  Not surprisingly, therefore, the idea of using spin 
clusters rather than single spins as the basic building block of 
exchange gate-based solid state quantum computation has attracted 
a great deal of recent theoretical attention.

In this Letter we point out and theoretically formulate 
an important conceptual, topological aspect of spin cluster 
qubits in solid state architectures.  In 
particular, we show that the proposed solid state two 
(or three)-dimensional spin cluster qubits will necessarily generate 
topological chiral terms in the qubit Hamiltonian in the presence of
external sources of flux, which will necessarily modify 
the Heisenberg exchange interaction bringing into question, 
in the process, the applicability of 
the exchange gate idea which has been the centerpiece
of spin-based solid state quantum computation architectures.  Thus the
proposed spin cluster qubits involving looped arrays of
localized, tunnel coupled electron spins will not 
behave as originally envisaged in 
the exchange gate scenarios discussed 
in many recent publications.  Even the proposed encoded spin qubits,
which specifically exclude all single qubit operations (and 
hence do not employ any applied external magnetic field), will have 
severe single-qubit decoherence problems (thus undermining 
the decoherence free properties of the logical qubit) 
since any temporally fluctuating
external sources of enclosed flux will cause
dephasing through the chiral term discussed here.
Contributions to the mean field component of an external 
magnetic field may arise, for example, from the nuclear 
dipolar field of the host lattice \cite{DeSousa}.  Alternatively, we 
find that the chiral 
term plays no role in clusters which do {\em not} contain three 
simultaneously, tunnel coupled sites.  The chiral term may also be considered 
negligible when the area enclosed by three tunneling channels
is small.  Atomic clusters, with small lattice spacings $\sim 1 $ \AA
$ $, require large external magnetic fields, $\sim 10^4$ T, to see a sizable 
chiral contribution.   We show that the large lattice spacing in 
looped quantum dot arrays enhances the role of three body chiral terms in the 
spin Hamiltonian. 
 
\begin{figure}
\includegraphics[width=2.3in]{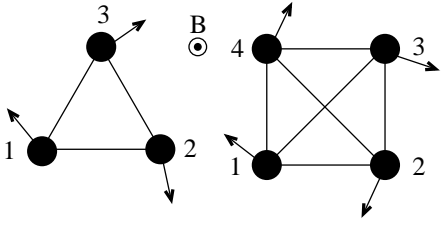}
\caption{ 
Two possible looped configurations of lateral  
quantum dots in a magnetic field.  
The triangular configuration contains three tunneling 
channels with one loop formed by the vertices $123$. 
The square configuration contains, in general, six tunneling 
channels with four three-site loops formed by the vertices $123$, $234$, $341$,
and $412$.    
\label{dots}}
\end{figure}

We study the low energy Hilbert space 
of $N$ lateral quantum dots containing $N$ electrons lying in the
$x-y$ plane by exact diagonalization of the following Hamiltonian:
\begin{eqnarray}
H= \sum_{i=1}^{N}\left[ \frac{{\textbf P}_i^2}{2m^*}
+V(\textbf r_i) \right] 
+\sum_{i<j}^{N} \frac{e^2}{\varepsilon \vert \textbf r_i-\textbf r_j \vert}
+H_Z, 
\label{FullH}
\end{eqnarray}
where we define the canonical 
momentum, $\textbf P=\textbf p+\frac{e}{c}\textbf
A$, and the Zeeman term $H_Z=g^*\mu_B \textbf S\cdot\textbf B$.  In 
GaAs we have an effective mass $m^*=0.067m_e$, dielectric
constant $\varepsilon=12.4$, and $g$-factor $g^*= -0.44$.  We 
work in the symmetric gauge with magnetic field $\bm{B}=B\hat{\bm{z}}$.  
The field couples directly to the total spin, $\textbf{S}$, through the 
Zeeman term.       
The single particle potential confines the electrons to lie at
the vertices of an equilateral triangle and square for $N=3$ and $4$, 
respectively, as shown in Fig.~\ref{dots}.  In diagonalizing the 
$N=3$ system we choose: 
$ V(\textbf r_i)=
(m^*\omega_0^2/2) \min_j\{\vert\textbf r_i-\textbf R_j\vert^2\}$,
where a parabolic confinement parameter, $\hbar\omega_0=6$ meV, 
effectively 
localizes the electrons at the sites $\bm R_j$ separated by $40$ nm.  

Prior to exact diagonalization 
of Eq.~(\ref{FullH}) we seek a perturbative expansion in terms of on-site spin operators.  
We consider the single band, tight binding limit, a good
approximation in the limit that the excited states of the quantum dot 
lie higher in energy than the lowest spin split states.  
We also take the on-site Coulomb interaction to be much larger 
than the tunneling energy.  (We verify  
numerically that, for the $N=3$ system, the following approximations 
are self consistent {\em only} at low $B$, $\omega_c \lesssim \omega_0/3 $,
where $\omega_c=eB/m^*c$ is the cyclotron frequency.)   
As a result we obtain the Hubbard Hamiltonian:
\begin{eqnarray}
H_{\text H}=
-\sum_{i,j,\alpha}t_{ij}c_{i\alpha}^{\dagger} c_{j\alpha}^{\vphantom{\dagger}}
+\sum_{i}\left(Un_{i\uparrow}n_{i\downarrow}+g^*\mu_B\textbf B \cdot \textbf S_i\right), 
\end{eqnarray}
where $c_{i\alpha}^{\dagger}$ creates a fermion at the site $i$ with
spin $\alpha\in\uparrow\downarrow$ and 
$n_{i\alpha}=c_{i\alpha}^{\dagger} c_{i\alpha}^{\vphantom{\dagger}}$.
We have defined the spin operators
$\textbf{S}_i=\frac{1}{2}c_{i\alpha}^{\dagger} \bm{\sigma}_{\alpha\alpha'} 
c_{i\alpha'}^{\vphantom{\dagger}}$ where $\bm{\sigma}$ are the Pauli
matrices.  As a consequence of the magnetic field the tunneling coefficients 
are complex:
$t_{ij}=\vert t_{ij} \vert \exp (2\pi i\Phi_{ij}/\Phi_0)$,
where $\Phi_0\equiv hc/e$.
The magnetic  
vector potential generates the Peierls phase:
$\Phi_{ij}=\int_{i}^{j}\textbf A\cdot d\textbf r$.
The integral is taken along a path connecting the sites 
$i$ and $j$.
Working in the small $\vert t_{ij}\vert/U$ 
limit we confine our attention to single occupancy states.  
Here we expand $H_{\text{H}}$ \cite{Macdonald} 
by applying a unitary transformation 
$\exp(iK)H_H\exp(-iK)$, where $K$ is an operator 
changing the number of doubly occupied states. 
Up to third order in this expansion we find\cite{Rokhsar, Sen}:
\begin{eqnarray}
&H_{\text{eff}}&= g^*\mu_B \textbf B \cdot\sum_{i}\textbf S_i
+\sum_{i,j}J_{ij} \textbf{S}_i\cdot \textbf{S}_j
\nonumber
\\
&+&\sum_{ijk\in\triangle}\mu_{ijk}
\textbf{S}_i\cdot(\textbf{S}_j\times\textbf{S}_k)
+\vartheta\left(\frac{t^4}{U^3}\right),
\label{eff}
\end{eqnarray}
where $t \sim \vert t_{ij} \vert$. 
The second term is the usual
Heisenberg interaction, where $J_{ij} = 2\vert t_{ij}\vert^2/U$, 
while the third term is a three site sum over chiral \cite{Wen} terms 
$\chi_{ijk}\equiv\mu_{ijk}
\textbf{S}_i\cdot(\textbf{S}_j\times\textbf{S}_k)$
around loops, $\triangle$. 
In $H_{\text{eff}}$ we have excluded fourth order terms of the 
form $(\textbf{S}_i\times\textbf{S}_j)\cdot(\textbf{S}_k\times\textbf{S}_l)$
and $(\textbf{S}_i\cdot\textbf{S}_j)(\textbf{S}_k\cdot\textbf{S}_l)$.
It was shown that the later plays a role at $B=0$ in a 
four spin, tetrahedral configuration \cite{Mizel}.
The coefficients in the chiral term:
$\mu_{ijk}= (24/U^2)\vert t_{ij}\vert \vert t_{jk}\vert \vert
t_{ki}\vert\text{sin}\left(2\pi\Phi_{ijk}/\Phi_0\right)$,
depend on the flux enclosed by the three site loop, $\Phi_{ijk}$. 
As a result, the chiral term generates an 
energy splitting between third order, 
virtual tunneling processes which run along and 
counter to the applied vector potential.  The phase, $2\pi\Phi_{ijk} /\Phi_0$, 
is the Aharonov-Bohm phase 
generated by the virtual current moving around the 
flux enclosed by the three site loop.
The chiral term is Hermitian but breaks time reversal symmetry and 
vanishes on bipartite lattices as a result of particle hole symmetry 
\cite{Sen}. 

We may now ask whether or not three coupled quantum 
dots containing three electrons may support a 
noiseless subsystem in the presence of a
fluctuating, perpendicular magnetic field \cite{Viola}.
The set of states comprising a noiseless subsystem 
remain invariant under the application of a suitable noise operator.  
We may, for example, choose the first term in Eq.~(\ref{eff}) as a
noise source.  
We then find the simplest example of a quantum dot, noiseless subsystem 
in the $S=1/2$ sector of the $N=3$ system.  
This encoding makes use of a fourfold, $B=0$ degeneracy 
found in this sector to protect quantum information stored in the
qubit defined by $\vert\lambda\rangle_{3}$, where $\lambda=0$ or $1$.  
The subscript denotes the number of quantum dots and electrons used to define the
qubit.  The four states may be written:

\begin{eqnarray}
\vert\lambda\rangle_{3} &\otimes& \vert -\frac{1}{2}\rangle = \frac{-1}{\sqrt{3}}
\left( \vert \downarrow \downarrow \uparrow \rangle 
+\omega^{\lambda +1}\vert\downarrow\uparrow\downarrow\rangle
+\omega^{2-\lambda }\vert\uparrow\downarrow\downarrow\rangle\right)  
\nonumber
\\
\vert\lambda\rangle_{3}&\otimes& \vert \frac{1}{2}\rangle
= \frac{1}{\sqrt{3}}
\left( \vert \uparrow \uparrow \downarrow \rangle 
+\omega^{\lambda +1}\vert\uparrow\downarrow\uparrow\rangle
+\omega^{2-\lambda }\vert\downarrow\uparrow\uparrow\rangle\right)  
\label{states}
\end{eqnarray}
where $\omega\equiv \exp(2\pi i/3)$.  
In the 
presence of a fluctuating Zeeman energy the first term in the 
tensor product preserves 
the quantum information stored in the quantum number $\lambda$ while 
the $z$-component of spin, the second term in the 
tensor product, may fluctuate.  
Application of $H_{\text{eff}}$ up to second order shows that the 
magnetic field only weakly affects $\vert\lambda\rangle_{3}$ 
through the Heisenberg term.
In fact several proposals \cite{Bacon,DiVincenzo,Lidar2,Levy,Meier}, 
suggest use of the  
Heisenberg term, with anisotropic couplings $J_{ij}$, 
to implement Pauli gates on encoded qubits. In Heisenberg 
gating schemes the expectation value of the individual Heisenberg
terms, and hence gates formed from them, remain insensitive to 
fluctuations in the Zeeman term because 
$\left[\bm{S}_i\cdot\bm{S}_j,\sum_i\bm{S}_i\right]=0$ for all $i$ and $j$.   

Up to second order, $H_{\text{eff}}$ also allows an  
encoding against Zeeman-like or collective noise.  
Here collective noise implies symmetry among all spins 
when coupling to a bath.  However, 
the third order chiral term acts as a {\em non-collective} 
system-bath interaction between the encoded qubit and 
a potentially noisy source of enclosed flux.  The chiral and Zeeman terms remove 
all degeneracies required to establish a qubit encoding immune to fluctuations
in the perpendicular magnetic field.    
Explicitly:
$ \chi_{123}\vert\lambda\rangle_{3}\otimes\vert \pm \frac{1}{2} \rangle 
=\mu_{123}(2\lambda-1)(\sqrt{3}/4)\vert\lambda\rangle_{3}\otimes\vert
\pm \frac{1}{2} \rangle$,
where $\mu_{123}=(12tJ/U)
\text{sin}\left(2\pi\Phi_{123}/\Phi_0\right)$ in the case $J_{ij}=J$
and $\vert t_{ij} \vert=t$ for all $i$ and $j$.
Therefore, $\chi_{123}$ yields an {\em effective} Zeeman splitting  
between the encoded basis states of the three spin qubit.  
The size of the 
splitting depends on $6 J (J/t)$ which, under reasonable 
conditions, cannot be neglected as long as the second order 
term, $J_{ij}\bm{S}_{i}\cdot\bm{S}_{j}$, remains 
large.  Gates constructed from the 
exchange interaction must operate on time scales shorter than 
typical spin relaxation times supplying, in turn, a 
necessary minimum to the exchange interaction.    
 
To test the accuracy of $H_\text{eff}$ we diagonalize 
Eq.~(\ref{FullH})
with $N=3$.  We construct the matrix
representing $H$ in the Fock-Darwin \cite{FockDarwin} basis 
centered in the triangle formed by the three 
parabola centers defined by $V(\textbf r)$.  We find 
it necessary to include up to $\sim 10^5$ origin centered, Fock-Darwin 
basis states with $z$-component of angular momentum 
less than twelve to obtain convergence.  We use 
a modified Lanczos routine to obtain the ground and excited states.      
We focus on the three lowest energy states in the absence of the 
Zeeman term.  
\begin{figure}
\includegraphics[width=2.5in]{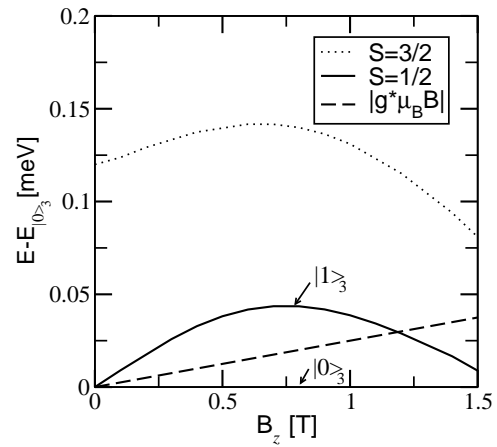}
\caption{ 
The energy of the lowest states of three electrons
confined in three adjacent, parabolic quantum dots obtained from exact 
diagonalization of Eq.~(\ref{FullH}).  The 
dot centers form an equilateral triangle in the $x-y$ plane 
with $40$ nm side lengths.  
The energies are plotted as a function of
perpendicular magnetic field.  The total $z$-component of 
spin is taken to be $1/2$.  
The energy of the state $\vert 0 \rangle_3$ in Eq.~(\ref{states}) is set to zero.  The 
solid and dotted lines show the energy of the states with total spin, 
$S=1/2$ and $S=3/2$, respectively.   The dashed line plots the Zeeman 
splitting in GaAs, $\approx 0.025 B[T]$ meV, for comparison.  
The energy splitting between the three-electron 
encoded qubit states $\vert 0\rangle_3$ and $\vert 1\rangle_3$ 
is larger than the Zeeman splitting between the spin up and down
states of a single electron.     
\label{graph}}
\end{figure}
Fig.~\ref{graph} shows the energy of the lowest states obtained
from exact diagonalization of $H$ in the $S^z=1/2$ sector as a
function of magnetic field.  The ground
state energy is set to zero.  The two lowest energy states have 
total spin $S=1/2$.  They are degenerate at $B=0$ as expected from the 
reflection symmetry of the triangular confining potential.  The 
next highest state has $S=3/2$ which corresponds to $6t^2/U\approx 0.125$ meV.  
Above this state we find (not shown) the higher excited states 
to lie near $3.8$ meV.  

As we increase magnetic field the
magnetic vector potential breaks the 
reflection symmetry of the confining potential 
leading to a splitting between the two
lowest states.   The splitting is linear in $B$, for small $B$, 
which suggests that the splitting 
is due to $\chi_{123}$.  The chiral 
term annihilates $S=3/2$ states.  We therefore expect that the energy of
the $S=3/2$
state, $E_{S=3/2}$, does not change with magnetic 
field at low fields (while  
$E_{S=3/2}-E_{\vert0\rangle_{3}}$
should increase linearly).  Here the relevant length scale is a
modified magnetic length,
$a=\sqrt{\hbar c/eB}(1+4\omega_0^2/\omega_c^2)^{-\frac{1}{4}}$.  As a result 
the magnetic field increases the Coulomb energy 
$(\sim e^2/ \varepsilon a)$
only above $B\sim 0.7$ T.  
At fields above $B\sim 0.7$ T the energies slope down indicating 
an eventual sign change in $J$, as for double quantum dots  
\cite{Loss2}.  Here the chiral contribution starts to become suppressed, along with
the exchange interaction.  
At larger fields the long range part of the Coulomb  
interaction becomes relevant.  
We must then keep extended terms in $H_\text{H}$ of the form 
$V\sum_{i,(\alpha,\alpha')\in\uparrow\downarrow}n_{i,\alpha}n_{i+1,\alpha'}$.
We note that inclusion of these terms into $H_{\text{H}}$ 
merely renormalizes the on-site contact term $U$ for $U\gg V$ \cite{Dongen}.  
For large $V$, the extended Hubbard term favors double occupancy of the 
dots and eventually leads to a sign change in $J$ \cite{Loss2}.  This 
suggests that $H_{\text{eff}}$ is qualitatively accurate below $B\sim 0.7$ T.
 
The slope of the energy splitting between the two lowest states in 
Fig.~\ref{graph} allows us to estimate $t/U$ for this system.  Using 
the area of the equilateral 
triangle defined by the dot centers we obtain $t/U\approx 0.09$  
which shows that our expansion in  $t/U$ is consistent at low 
fields \cite{corr}.  For $N=3$, 
only odd powers of $t_{ij}$ allow linear magnetic 
field dependence in the splitting, showing
that the magnetic field dependence captured by the chiral term 
is accurate up to $\vartheta\left(\frac{t^5}{U^4}\right)$.

The dashed line in Fig.~\ref{graph} shows, for comparison, the Zeeman  
splitting between the spin states of a single electron.  The 
Zeeman splitting is smaller than the energy difference 
$E_{\vert 1\rangle_3}- E_{\vert 0\rangle_3}$ suggesting that, in the 
chosen parameter regime,  
quantum information encoded in the 
``noiseless'' spin states of a three electron-triangular set of   
quantum dots is {\em more} sensitive to a perpendicular magnetic field 
than a single electron.

We now study magnetic field effects on a 
decoherence free subspace (DFS) \cite{Whaley} formed 
from quantum dot loops.  
The lowest number of physical  
spins supporting a DFS is four.  For simplicity,  we begin with 
four quantum dots containing four electrons 
lying at the vertices of a square 
with equal tunneling $\vert t_{ij} \vert= t$, including diagonal terms shown 
schematically in Fig.~\ref{dots}.   
In this case we find a DFS among two $S=0$ states corresponding to 
$\lambda=0$ and $1$:
\begin{eqnarray}
&&\vert\lambda\rangle_{4}=\vert\uparrow\uparrow\downarrow\downarrow\rangle
+\vert\downarrow\downarrow\uparrow\uparrow\rangle
+\omega^{\lambda+1}\vert\uparrow\downarrow\uparrow\downarrow\rangle
\nonumber
\\
&&+\omega^{\lambda+1}\vert\downarrow\uparrow\downarrow\uparrow\rangle
+\omega^{2-\lambda}\vert\downarrow\uparrow\uparrow\downarrow\rangle
+\omega^{2-\lambda}\vert\uparrow\downarrow\downarrow\uparrow\rangle.
\end{eqnarray}
These states have equal Zeeman energies and, up to second order 
in Eq.~(\ref{eff}), show no direct magnetic field dependence 
in their energy spectra.  
As for the third order term the spin Hamiltonian must respect 
the exchange symmetry inherent in the lattice.  Therefore, the 
sum over asymmetric chiral terms must vanish with equal tunneling
among all sites.  In the $\vert\lambda\rangle_{4}$ basis we find:  
\begin{eqnarray}
\sum_{ijk\in\triangle} \chi_{ijk}\otimes I_l \vert\lambda\rangle_4
=\frac{\sqrt{3}}{4}(2\lambda-1)\sum_{ijk\in\triangle}\mu_{ijk}
\epsilon_{ijkl}
\vert\lambda\rangle_4,
\end{eqnarray}
where $\epsilon_{ijkl}$ is the four component Levi-Civita symbol 
and $I_l$ is the identity operator.  The sums run over three site
loops excluding $l= i,j$ or $k$.  
The sum vanishes with tunneling $\vert t_{ij}\vert
=t$ for all $i$ and $j$.  
However, if we impose some asymmetry 
in tunneling, as is done when applying a Pauli gate to the encoded
qubit, this sum is not zero.  Consider a simple case:
$\vert t_{31} \vert=\vert t_{23} \vert = \vert t_{34} \vert 
= t(1+\delta) $, 
where $\delta$ is a number and all other $\vert t_{ij} \vert=t$.   
The sum over chiral terms then induces an energy splitting 
$\simeq 24\pi\sqrt{3} t J\delta A B_z /(U \Phi_0)$ between 
the states with $\lambda=0$ and $1$ for $\Phi_{ijk}/\Phi_0\ll 1$.  
Here $A$ is the area of the triangle defined by the vertices $123$.      
As in the $N=3$ system, the energy splitting acts as an effective Zeeman 
splitting on the $N=4$ encoded qubit with the exception 
that here the parameters $A$ {\em and} $\delta$ may be included in an  
effective $g$-factor of the encoded two-level system.   

As mentioned earlier, Pauli gating sequences may be applied to
encoded, multi-spin qubits by tuning the Heisenberg 
couplings, and therefore $t_{ij}$, in an asymmetric fashion.  
When applied to a DFS a Pauli gate composed of Heisenberg terms must, by construction, involve 
a spin specific asymmetry with, for example, large $\delta$.
As shown above, the $N=4$ encoded qubit will, during a 
gate pulse, be sensitive to sources of enclosed flux because the scalars 
forming the Heisenberg interaction do not 
commute with the individual chiral terms where, for example: 
$\left[\bm{S}_i\cdot\bm{S}_3,\chi_{123}\right]\neq 0$, with 
$i=1$ or $2$.  

We have shown that, in multiple quantum dot architectures  
containing tunnel coupled loops, fluctuations in enclosed flux provide a potential 
source of phase flip error in noiseless subsystems and subspaces through 
chiral currents generated by virtual hopping processes 
around three site loops.  In systems for which fluctuations in the 
phase of $t_{ij}$ can be ignored we have shown that 
a well controlled source of flux  
may also be a good candidate for implementing a logical Pauli $Z$ gate 
on a three or four spin encoded qubit.  
Quantum algorithms using only exchange based quantum gates require
several, accurate applications of the exchange Hamiltonian
\cite{DiVincenzo}.  The chiral term may then supplement
the exchange interaction and therefore potentially reduce the overhead
in exchange based algorithms.
However, the phase in $t_{ij}$ is in general a Berry's 
phase \cite{Berry} arising from  
symmetry breaking terms in the original Hamiltonian.  
Spin orbit coupling can contribute to Berry's phase effects in triangular lattices 
\cite{Lyanda-Geller} and may therefore play a role in effective spin Hamiltonians 
modeling tunnel coupled quantum dots.


This work is supported by ARDA and NSA-LPS.




\begin{thebibliography}{}

\bibitem{Loss} D. Loss and D. P. DiVincenzo, Phys. Rev. A 
\textbf{57}, 120 (1998). 

\bibitem{Kane} B.E. Kane, 
Nature  \textbf{393}, 133 (1998).

\bibitem{Hu} X. Hu and S. Das Sarma,
Phys. Rev. A \textbf{61}, 062301 (2000);  
X. Hu and S. Das Sarma, 
Phys. Rev. A \textbf{64}, 042312 (2001).

\bibitem{Vrijen} R. Vrijen {\em et al.},
Phys. Rev. A \textbf{62}, 012306 (2000).

\bibitem{DeSousa} A. V. Khaetskii and Y. V. Nazarov,
Phys. Rev. B \textbf{64}, 125316 (2001); 
R. de Sousa and S. Das Sarma,
Phys. Rev. B \textbf{67}, 033301 (2003).


\bibitem{Hu2} X. Hu {\em et al.},
Phys. Rev. Lett. \textbf{86}, 918 (2001).

\bibitem{DiVincenzo} D.P. DiVincenzo {\em et al.},
Nature \textbf{408}, 339 (2000).

\bibitem{Benjamin} S. C. Benjamin and S. Bose,
Phys. Rev. Lett. \textbf{90}, 247901 (2003).

\bibitem{Levy} J. Levy, 
Phys. Rev. Lett. \textbf{89}, 147902 (2002).

\bibitem{Meier} F. Meier {\em et al.},
Phys. Rev. Lett. \textbf{90}, 047901 (2003);  
F. Meier {\em et al.}, Phys. Rev. B \textbf{68}, 134417 (2003).


\bibitem{Bacon} D. Bacon {\em et al.},
Phys. Rev. Lett. \textbf{85}, 1758 (2000).

\bibitem{Viola2} L. Viola {\em et al.},
Phys. Rev. Lett. \textbf{85}, 3520 (2000).


\bibitem{Kempe}  J. Kempe {\em et al.},
Phys. Rev. A \textbf{63}, 042307 (2001).

\bibitem{Whaley} D. A. Lidar {\em et al.},
Phys. Rev. Lett. \textbf{81}, 2594 (1998).  

\bibitem{Viola} E. Knill {\em et al.},
Phys. Rev. Lett. \textbf{84}, 2525 (2000).


\bibitem{Macdonald} M. Takahashi, 
J. Phys. C \textbf{10}, 1289 (1977);
A. H. MacDonald {\em et al.},
Phys. Rev. B \textbf{37}, 9753 (1988).  

\bibitem{Rokhsar} D. S. Rokhsar,
Phys. Rev. Lett. \textbf{65}, 1506 (1990). 

\bibitem{Sen} D. Sen and R. Chitra,
Phys. Rev. B \textbf{51}, 1922 (1995).  

\bibitem{Wen} X. G. Wen {\em et al.}, 
Phys. Rev. B \textbf{39}, 11413 (1989). 

\bibitem{Mizel} A. Mizel and D. A. Lidar,
Phys. Rev. Lett. \textbf{92}, 077903 (2004).

\bibitem{Lidar2} M. S. Byrd and D. A. Lidar,
Phys. Rev. Lett. \textbf{89}, 047901 (2002).

\bibitem{FockDarwin} V. Fock, Z. Phys. \textbf{47}, 446 (1928);  
C.G. Darwin, Proc. Cambridge Philos. Soc. Math. Phys. Sci. 
\textbf{27}, 86 (1930).

\bibitem{Loss2} G. Burkard {\em et al.},
Phys. Rev. B \textbf{59}, 2070 (1999). 

\bibitem{Dongen} P. G. J. van Dongen, 
Phys. Rev. B \textbf{49}, 7904 (1994).

\bibitem{corr} The effective enclosed area includes a correction 
due to the parabolic confinement.  V.W. Scarola and S. Das Sarma 
(to be published).
 
\bibitem{Berry} M. V. Berry, 
Proc. Roy. Soc. London, Ser. A \textbf{392}, 45 (1984).
 
\bibitem{Lyanda-Geller} Y. Lyanda-Geller {\em et al.},
Phys. Rev. B \textbf{63}, 184426 (2001). 

\end{thebibliography}
\end{document}